\documentclass[epj,twocolumn]{webofc}
\usepackage[varg]{txfonts}   
\usepackage{subfigure}
\usepackage{graphics}
\graphicspath{%
               {./}
}%

\woctitle{Advances in Radioactive Isotope Science}
\begin{document}

\selectlanguage{english}

\title{Study of medium-mass and heavy hypernuclei produced through spallation and fission reactions in inverse kinematics}
%
%

\author{
J. L. Rodr\'{i}guez-S\'{a}nchez\inst{1}\thanks{\emph{Corresponding author: j.l.rodriguez.sanchez@udc.es}},
\and
J. Cugnon\inst{2}
\and 
J.-C. David\inst{3}
\and
J. Hirtz\inst{3,4}
\and
A. Keli\'{c}-Heil\inst{5}
}

\institute{CITENI, Campus Industrial de Ferrol, Universidade da Coru\~{n}a, E-15403 Ferrol, Spain
\and 
AGO department, University of Li\`{e}ge, all\'{e}e du 6 ao\^{u}t 19, b\^{a}t.~B5, B-4000 Li\`{e}ge, Belgium
\and 
IRFU, CEA, Universit\'{e} Paris-Saclay, F-91191 Gif-sur-Yvette, France
\and 
Physics Institute, University of Bern, Sidlerstrasse 5, 3012 Bern, Switzerland
\and 
GSI Helmholtzzentrum f\"{u}r Schwerionenforschung, Planckstra\ss e 1, D-64291 Darmstadt, Germany
}

\abstract{%
Innovative experiments using the inverse kinematics technique to accelerate light, medium-mass, and heavy nuclei at relativistic energies have become excellent tools to produce and study hypernuclei. In this work, we investigate hypernuclei created in spallation reactions, where multifragmentation, particle evaporation, and fission processes play an important role in the formation of final hypernuclei residues. For the description of spallation reactions, we couple the Li\`{e}ge intranuclear cascade model, extended recently to the strange sector, to a new version of the ablation (ABLA) model that accounts for the evaporation of $\Lambda$-particles from hot hyperremnants produced during the intranuclear cascade stage. These state-of-the art models are then used to study the production of hypernuclei close to the drip lines through spallation-evaporation and fission reactions. Moreover, recent results obtained for the study of hypernuclei dynamics, in particular, for the constraint of the viscosity parameter involved in hyperfission reactions are also presented.
}
\maketitle
\section{Introduction}

Since the discovery in 1952 of the first hypernucleus in an experiment carried out with emulsion chambers~\cite{Danysz1953}, much effort has been invested in extending our knowledge of the nuclear chart towards the SU(3) flavor octects~\cite{Lenske2018}. Hypernuclei are bound nuclear systems of strange baryons, the so-called hyperons, such as $\Lambda$, $\Sigma$, $\Xi$, or $\Omega$, produced in high energy collisions. These hyperons can be captured by nuclei since their life times are longer than the characteristic reaction times ($\sim10^{-23}$~s)~\cite{Park2000}. Consequently, the study of hypernuclei and their properties provides the opportunity to provide information on the strange-matter properties and on the hyperon-nucleon (YN) and hyperon-hyperon (YY) interactions \cite{Millener1988,Maessen1989,Tominaga98,Cugnon2000,Schulze2010,Hiyama2018,Haidenbauer2020}, which cannot be determined from scattering experiments.

Moreover, the investigation of hypernuclear structure has important implications in the study of compact astrophysical objects. As shown in several works~\cite{Vidana2000,Schulze2011,Bedaque2015,Wirth2016,Fortin2017},  
the composition and equation of state (EOS) of supernovae and neutron star (NS) cores are poorly known due to their dependence on the hyperon content, being the hypernuclear weak decay the only available tool to acquire knowledge on strangeness-changing weak baryon interactions. The understanding of the weak interactions, as well as the strong ones, have a direct connection with astrophysics since they are important inputs when investigating the composition and macroscopic properties (masses and radii) of compact stars, their thermal evolution, and stability. The presence of hyperons in finite and infinite nuclear systems, such as hypernuclei and neutron stars, also make it necessary to extend the investigation of the nuclear dissipation mechanisms to the strangeness sector because they play a crucial role in the understanding of the  oscillation modes of NSs. Significant effort has been aimed at understanding whether gravitational-wave emission sets the upper rotational frequency limit for pulsars, e.g., via the r-mode instability discovered by Andersson, Friedman, and Morsink~\cite{Andersson1998,Friedman1998} in 1998. This possibility is of particular interest since r-modes can lead to the emission of detectable gravitational waves in hot and rapidly rotating NSs. R-modes are predominantly toroidal oscillations (i.e., oscillations with a divergenceless velocity field and a suppressed radial component of the velocity) of rotating stars restored by the Coriolis force~\cite{Kolomeitsev2015}, acting similar to Rossby waves in Earth’s atmosphere and oceans. It is of great importance to understand whether internal fluid dissipation allows the instability to develop in such systems or whether it suppresses the r-modes completely.

\section{Theoretical framework}

The collision between the proton and target nuclei, the so-called spallation reaction, is described with the latest C++ version of the dynamical Li\`{e}ge intranuclear-cascade model (INCL)~\cite{Mancusi14} coupled to the ablation model ABLA~\cite{JL2022}, which are based on Monte Carlo techniques obeying all conservation laws throughout each reaction event. INCL describes the spallation reaction as a sequence of binary collisions between the nucleons (hadrons) present in the system. Nucleons move along straight trajectories until they undergo a collision with another nucleon or until they reach the surface, where they could possibly escape. The latest version of the INCL also includes isospin- and energy-dependent nucleon potentials calculated according to optical models~\cite{Bou13}, as well as isospin-dependent pion potentials~\cite{Aoust2006}. INCL has been recently extended toward high energies ($\sim$20 GeV) including new interaction processes, such as multipion production~\cite{Mancusi2017}, production of $\eta$ and $\omega$ mesons~\cite{JCD2018}, and strange particles~\cite{Jason2018,JL2018,Jason2020}, such as kaons and hyperons. Therefore this new version of INCL allows us to predict the formation of hot hyperremnants and their characterization in atomic ($Z$) and mass ($A$) numbers, strangeness number, excitation energy, and angular momentum. We remark that the good agreement of INCL calculations with experimental kaon production cross sections obtained from proton-induced reactions on light, medium-mass, and heavy nuclei at energies of few GeV~\cite{Jason2020}, as well as the reasonable description of hypernuclei production cross sections through strangeness-exchange reactions ($\pi^{+}$,$K^{+}$)~\cite{JL2018}, allow us to guarantee a correct prediction of the excitation energy gained by the hyperremnants after the proton-nucleus collision.

The hyperremnants enter then the deexcitation stage that is modeled by the code ABLA~\cite{JL2022}. This model describes the deexcitation of a nuclear system through the emission of $\gamma$-rays, neutrons, $\Lambda$-hyperons, light-charged particles, and intermediate-mass fragments (IMFs) or fission decays in case of hot and heavy compound nuclei. The particle emission probabilities are calculated according to the Wei$\beta$kopf-Ewing formalism~\cite{We40}, being the separation energies and the emission barriers for charged particles calculated according to the atomic mass evaluation from 2016~\cite{mass2020} and the prescription given by Qu and collaborators~\cite{Qu2011}, respectively. The fission decay width is described by the Bohr-Wheeler transition-state model~\cite{BW39}, following the formulation given by Moretto and collaborators~\cite{Moretto75}. The slowing effects of nuclear dissipation are also included by using the Kramers approach~\cite{KR40} together with transient time effects~\cite{BJ03}.

\section{Results}

\begin{figure}[t!]
\centering
\subfigure{\label{fig:1a}\includegraphics[width=0.49\textwidth,keepaspectratio]{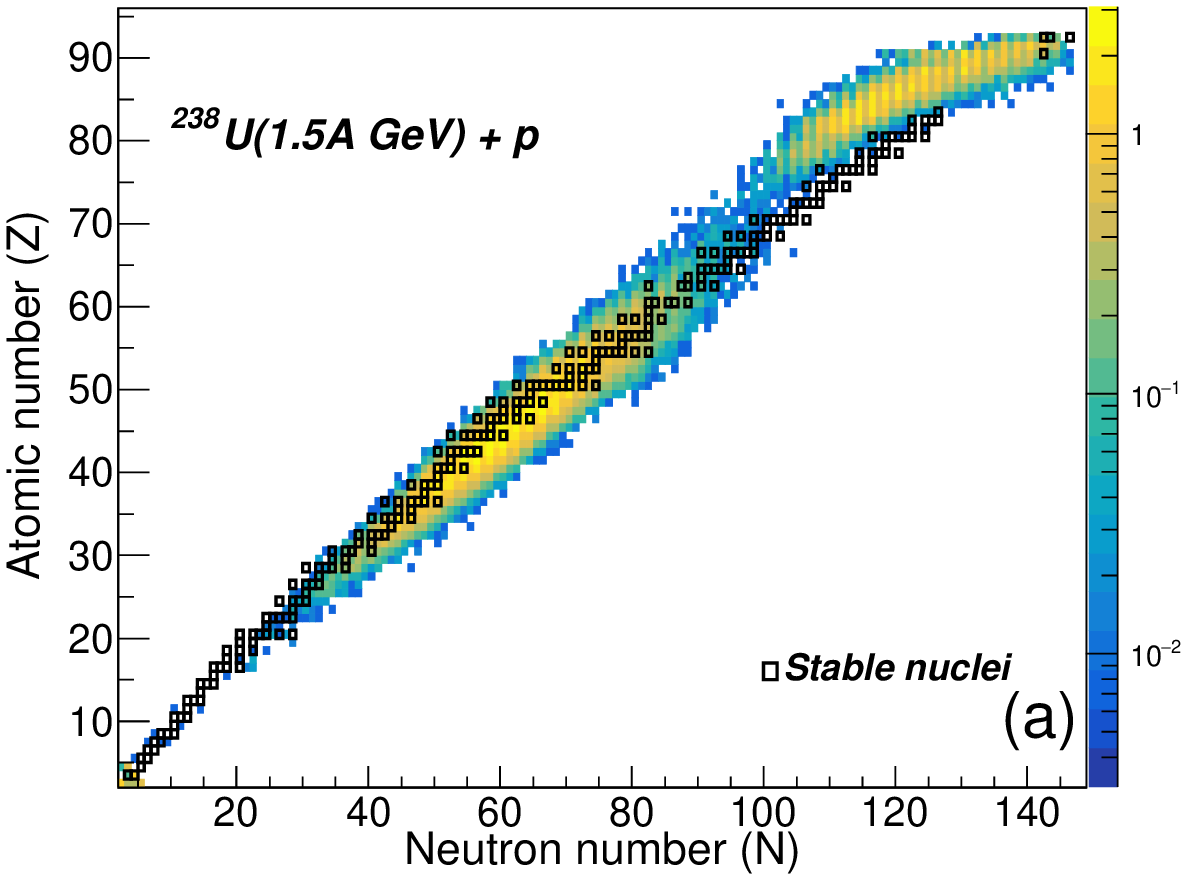}}
\centering
\subfigure{\label{fig:1b}\includegraphics[width=0.49\textwidth,keepaspectratio]{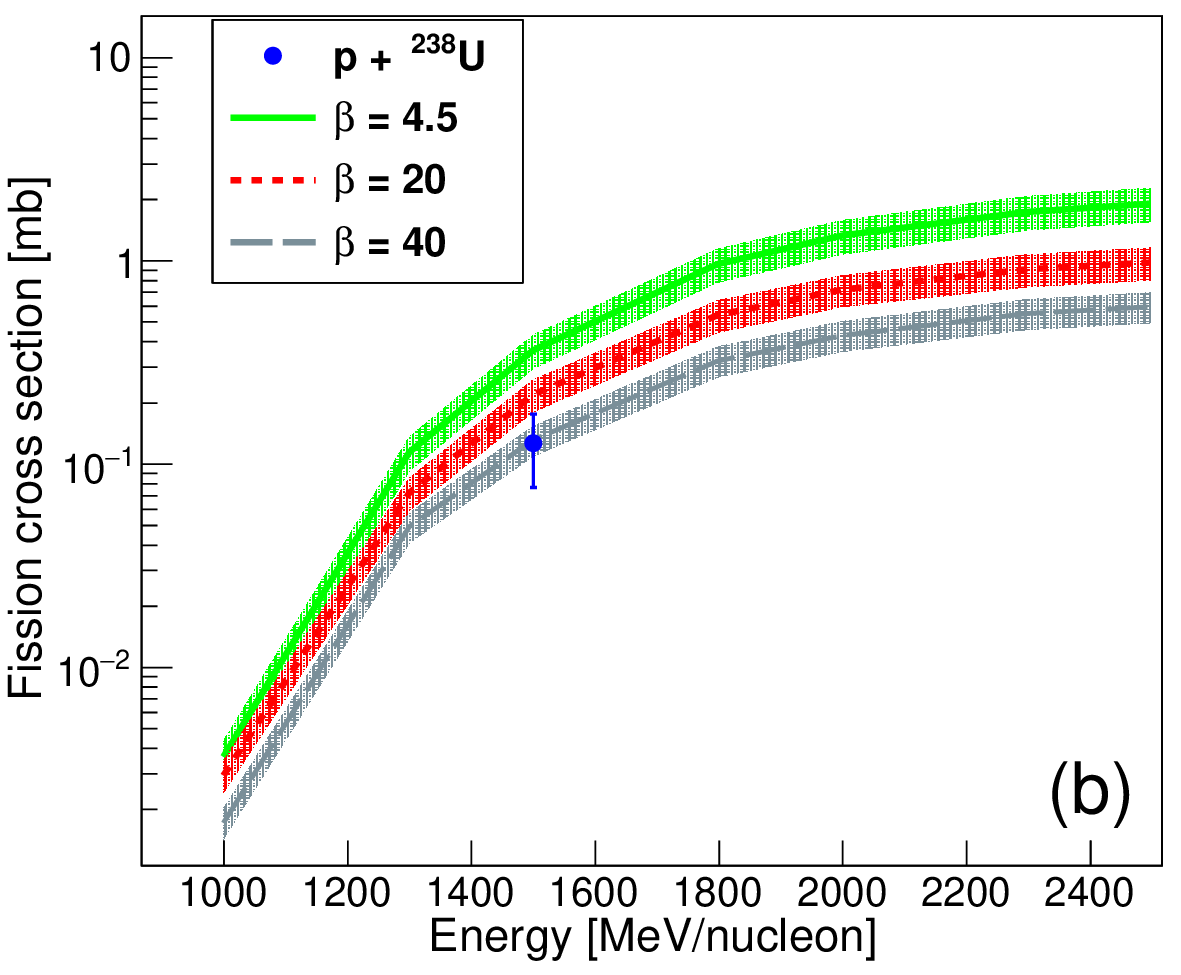}}
\caption{(Color online) (a) Two-dimensional cluster plot of the isotopic production cross sections of $\Lambda$-hypernuclei in $\mu$b shown as a chart of nuclides for proton-induced reactions on $^{238}$U at 1.5$A$ GeV. Open squares correspond to normal stable nuclei. (b) Hypernuclear fission cross section (dots) as a function of the projectile kinetic energy per nucleon for target nuclei of $^{238}$U. The lines correspond to dynamical fission calculations for different values of the dissipation coefficient $\beta$. Dashed areas represent the uncertainties.
}
\label{fig:12}
\end{figure}

Fig.~\ref{fig:1a} shows an example of the isotope composition of hypernuclei produced by evaporation, multifragmentation, and fission processes of $^{238}$U projectiles at a kinetic energy of 1.5$A$ GeV impinging onto a proton target. The production of hypernuclei is represented as a function of the proton and neutron number of each hyperfragment because this allows us for the best overview . For instance, heavy neutron-deficient hypernuclei ($70<Z<89$) close to the proton drip line can be investigated through spallation reactions while neutron-rich hypernuclei ($25<Z<60$) are more  through fission reactions.

In Fig.~\ref{fig:1a} the dissipation coefficient used to describe the fission process of hyperremnants was constrained according to the methodology shown in Fig.~\ref{fig:1b}, where we display the experimental cross section obtained for hypernuclear fission reactions induced in $^{238}$U ions~\cite{Ohm1997} as a function of the proton kinetic energy. The data is compared to INCL+ABLA calculations in which we have assumed different values for the dissipation parameter $\beta$: 4.5 (solid line), 20 (short-dashed line), and 40 (long-dashed line)~$ \times 10^{21}$ s$^{-1}$. We can see that the hypernuclear fission cross sections decrease when increasing the value of the viscosity parameter, which is expected since the nuclear system evolves more slowly, needing more time to reach the saddle point configuration. This fact favors the cooling of the nuclear system by particle emission reducing the fission probabilities. In these calculations we also take into account the uncertainties in the nuclear level densities and fission barrier heights, which are displayed in the figures with dashed areas. These uncertainties do not exceed 18$\%$ of the total hypernuclear fission cross section, being the 16$\%$ of this uncertainty attributed to the fission barrier height. The comparison allowed us to constrain the value of the viscosity coefficient resulting in an average value of $(28 \pm 12) \times 10^{21}$~s$^{-1}$, as detailed in Ref.~\cite{JL2023}.

\section{Conclusions}
The de-excitation model ABLA originally written in fortran has been translated to C++ and extended to the strangeness sector including the binding energies of $\Lambda$ particles and its propagation through decay processes such as particle evaporation, multifragmentation, and fission. This improvement opens the possibllity of investigating the formation of cold light, intermediate-mass, and heavy hypernuclei far from the stability region of normal nuclei. In particular, we have studied in this work the production of $\Lambda$-hypernuclei in spallation reactions by coupling the de-excitation model ABLA to the version 6.0 of the Li\`{e}ge intranuclear cascade model INCL, which has also been extended recently to the strangeness sector.

The hypernuclei fission cross sections measured at the COSY-J\"{u}lich facility with high energetic protons impinging onto a target nucleus of $^{238}$U was used for the first time to investigate the nuclear dissipation coefficient in the presence of hypernuclear matter. The experimental data was compared to our state-of-the art dynamical model. The comparison allowed us to constrain the dissipation parameter in fission of hypernuclear matter, resulting in an average value of $(28 \pm 12) \times 10^{21}$~s$^{-1}$~\cite{JL2023}. This finding is 6 times larger than that obtained for normal nuclear matter~\cite{Nadtochy2007,CS2007,JL2014,Rodriguez2016}, which implies that in presence of hyperons the conversion of intrinsic energy into collective motion goes much slower.

\section*{Acknowledgments}
J.L.R.-S. is thankful for the support from Xunta de Galicia under the postdoctoral fellowship Grant ED481D-2021-018 and from the "Ram\'{o}n y Cajal" program under the Grant RYC2021-031989-I, funded by MCIN/AEI/10.13039/501100011033 and by “European Union NextGenerationEU/PRTR”.

\end{document}